\documentclass[useAMS,usenatbib]{mn2e}
\usepackage{subfigure}
\usepackage{graphicx}
\usepackage{color}
\usepackage{soul}
\usepackage{amssymb, amsmath}
\usepackage{threeparttable}
\title[Sizes of Red and Blue Globular Clusters]{Is there a Size Difference
  between Red and Blue Globular Clusters?}
\author[J.~M.~B. Downing]{J. M. B. Downing$^{1}$\thanks{E-mail:
    downin@ari.uni-heidelberg.de}\\
  $^{1}$Astronomisches Rechen-Institue, Zentrum f\"{u}r Astronomie der
  Universit\"{a}t Heidelberg, M\"{o}nchhofstra\ss{}e 12-14, D-69120
  Heidelberg, Germany}
\begin{document}
\date{Accepted ... Received ... in original form ...}
\pagerange{\pageref{firstpage}--\pageref{lastpage}} \pubyear{2012}
\maketitle
\label{firstpage}
\begin{abstract}

  Blue (metal-poor) globular clusters are observed to have half-light radii
  that are $\sim$20\% larger than their red (metal-rich) counterparts.  The
  origin of this enhancement is not clear and differences in either the
  luminosity function or in the actual size of the clusters have been
  proposed.  I analyze a set of dynamically self-consistent Monte Carlo
  globular cluster simulations to determine the origin of this enhancement.  I
  find that my simulated blue clusters have larger half-light radii due to
  differences in the luminosity functions of metal-poor and metal-rich stars.
  I find that the blue clusters can also be physically larger, but only if
  they have a substantial number of black holes heating their central regions.
  In this case the difference between half-light radii is significantly larger
  than observed.  I conclude that the observed difference in half-light radii
  between red and blue globular clusters is due to differences in their
  luminosity functions and that half-light radius is not a reliable proxy for
  cluster size.

\end{abstract}

\section{Introduction}
\label{sec:Intro}

Globular cluster (GC) systems are common in disk and elliptical galaxies and
in the past decade significant observational progress has been made in
understanding them.  These globular clusters are frequently used to help
understand the formation and evolution of their host galaxies.  However, the
observational properties of globular clusters are not fully understood.  A
case in point is the half-light radius ($r_{hl}$).  Several models show that
the the half-light radius of a cluster should remain fairly constant during its
lifetime \citep{SpitzerThuan72, AarsethHeggie98} so it is often used to
compare the structures of globular clusters of different ages.  Several
observational studies, however, show that metal-poor (blue) globular clusters
have half-light radii that are systematically larger (by $\sim 20\%$) than
their metal-rich (red) counterparts \citep{KunduWhitmore98, KunduWhitmore01,
  KunduEtAl99, PuziaEtAl99, LarsenEtAl01a, LarsenEtAl01b, BarmbyEtAl02,
  HarrisEtAl02, JordanEtAl04, Harris09}.

It is not clear what the origin of this discrepancy is nor is it certain that
it truly represents a difference in the sizes of the clusters as measured by
their half-mass radii ($r_{hm}$).  \cite{LarsenBrodie03} proposed that the
observed enhancement in the size of the blue clusters could be due to a
projection effect.  They note both that blue and red globular clusters follow
different radial distributions and that in the Milky Way there is a
relationship between cluster size and galactocentric distance
\citep{VanDenBerghEtAl91}.  They argue that if such a relationship exists in
all galaxies, the differing radial distributions of red and blue clusters could
account for the observed difference in sizes.  A detailed survey by
\cite{Harris09}, however, showed that the ratio of $r_{hl}$ between blue and
red clusters does not depend on galactocentric distance, as would be predicted
by the \cite{LarsenBrodie03} model.  Furthermore, \cite{LarsenBrodie03} give
no reason why a galactocentric distance-globular cluster size trend should
exist in other galaxies.

\cite{Jordan04} explained the difference as a result of differing stellar
evolution processes in metal-poor and metal-rich stellar populations.  Metal
poor stars lose less mass and have longer main-sequence lifetimes than
metal-rich stars.  Assuming that the difference between red and blue stellar
populations is metallicity, this leads to differing luminosity functions in
blue and red GCs.  Using multi-mass Michie-King models with fixed half-mass
radii and a stellar population with an age of 13 Gyrs, \cite{Jordan04} was
able to re-produce the observed $r_{hl}$ enhancement in blue globular
clusters.  However, these models are not dynamical simulations of globular
cluster evolution and the results are valid only if $r_{hm}$ is the same in
red and blue globular clusters.

By contrast \cite{Schulman12} used direct $N$-body simulations to investigate
the differences in dynamical evolution between blue and red open clusters.
They also assumed that the colour of a cluster reflects its metallicity and
argued that, because metal-poor stars lose mass more slowly than metal-rich
stars, blue clusters will lose less mass to stellar evolution over their
lifetimes than will red clusters.  This reduces the gravitational potential of
red clusters, causing them to expand and become larger than blue clusters.
However, scattering interactions between the more massive metal-poor stars
will be more energetic than those between less massive metal-rich stars and
the blue clusters will experience stronger two-body heating.  This will cause
them to expand relative to the red clusters once the initial phase of rapid
mass-loss is concluded.  If the effect of two-body heating is stronger, blue
clusters will be physically larger than red clusters.  \cite{Schulman12}
showed that the half-mass radii of their simulated blue clusters were indeed
$\sim 20\%$ larger after several initial half-mass relaxation times ($t_{rh}$)
than the half-mass radii of their simulated red clusters.  They also found
little difference between the ratio of $r_{hm}$ and $r_{hl}$ between blue and
red clusters.  Therefore they conclude that the enhancement in $r_{hl}$ in
blue globular clusters is due to an actual size enhancement and has a
dynamical origin.  The \cite{Schulman12} models, however, contain 10-100 times
fewer stars than are present in globular clusters and consequently relax and
dissolve when they are only a few hundred Myr old, far younger than the 10-13
Gyr age of many globular cluster systems.  The relationship between stellar
evolution timescales and dynamical timescales is also quite different in
these simulations than in GCs.

In this paper I re-visit the problem of size differences between blue and red
globular clusters by analyzing a set of Monte Carlo star cluster models.
These are self-consistent dynamical simulations that contain a similar number
of stars to globular clusters, include parametrized stellar evolution and
primordial binaries.  I will investigate whether my  blue globular clusters
are larger than the red ones, as was reported by \cite{Schulman12}.  If they
are, I will determine if $r_{hl}$ is enhanced to  the same degree as $r_{hm}$
and if the processes reported by \cite{Jordan04} have an significant effect.
I will also be able to determine weather or not $r_{hl}$ is a good
observational proxy for $r_{hm}$.

\section{Numerical models}
\label{sec:Models} 

\begin{table}
  \caption[Simulation Parameters]{Parameters of the simulations.  The first
    column gives the identifying label, the second gives the metallicity, the
    third the total mass of the cluster, the fourth the initial half-mass
    radius and the fifth the initial half-mass relaxation
    time.\label{tab:SimParams}}
  \begin{tabular}[c]{l r r r r}
    \hline
    \multicolumn{5}{c}{Simulation Parameters}\\
    \hline
    Simulation & $Z$ & M [M$_{\odot}$] & $r_{hm}$ [pc] &
    $t_{rh}$ [Myr] \\
    \hline
    red21 & 0.02 & $3.61 \times 10^{5}$ & 7.14 & $3.54 \times 10^{3}$ \\
    red37 & 0.02 & $3.63 \times 10^{5}$ & 4.05 & $1.51 \times 10^{3}$ \\
    red75 & 0.02 & $3.62 \times 10^{5}$ & 2.00 & $5.25 \times 10^{2}$ \\
    blue21 & 0.001 & $3.60 \times 10^{5}$ & 7.14 & $3.55 \times 10^{3}$ \\
    blue37 & 0.001 & $3.62 \times 10^{5}$ & 4.05 & $1.51 \times 10^{3}$ \\
    blue75 & 0.001 & $3.62 \times 10^{5}$ & 2.00 & $5.25 \times 10^{2}$ \\
    \hline
  \end{tabular}
\end{table}

The simulations in this paper were performed using a Monte Carlo method to
self-consistently simulate the dynamical evolution of a star cluster in the
Fokker-Planck two-body relaxation limit.  The primary advantage to this method
over direct $N$-body is speed; the Monte Carlo code scales with
$\mathcal{O}(N^{1}) - \mathcal{O}(N^{2})$, where $N$ is the number of stars in
the system, as opposed to $\mathcal{O}(N^{3}) - \mathcal{O}(N^{4})$ for direct
$N$-body codes.  Thus it it is possible to run globular cluster-sized
simulations over a full Hubble time of dynamical evolution, a task currently
impossible to do using a direct $N$-body code.  Unlike standard Fokker-Planck
codes, however, the Monte Carlo method treats the cluster as an ensemble of
particles and can provide the same star-by-star information as a direct
$N$-body simulation.  This also makes it much simple to include prescriptions
for stellar evolution and strong few-body interactions.  The Monte Carlo code
is described in detail by  \cite{Giersz98, Giersz01, Giersz06} while the
strong few-body interactions are calculated as described in
\cite{GierszSpurzem03}.  The code computes parametrized evolution for each
star and binary using the SSE \citep{HurleyEtAl00} and BSE
\citep{HurleyEtAl02} stellar evolution prescriptions \cite{GierszEtAl08}.
These are essentially the same prescriptions as used by \cite{Schulman12}.  The
code has been compared to, and provides excellent agreement with, both direct
N-body simulations and observations \citep{GierszEtAl08, HeggieGiersz08,
  GierszHeggie09, HeggieGiersz09}.

The simulations used in this paper are a subset of a collection of simulations
that were originally performed to investigate the dynamical creation of black
hole binaries in globular clusters.  They are described in some detail in
\cite{DowningEtAl10}.  All simulations are initialized as $5 \times 10^{5}$
particle Plummer spheres with 10\% primordial binaries.  Initial binary
parameters are chosen according to the eigenvalue evolution and feeding
algorithms of \cite{Kroupa95}.  The initial stellar density is controlled by
the ratio of the tidal radius ($r_{t}$) to $r_{hm}$.  For all simulations
$r_{t} = 150$ pc and I use $r_{t}/r_{hm} \in \lbrace 21, 37, 75 \rbrace$,
corresponding to initial stellar number densities within $r_{hm}$ of $\sim
10^{2}$, $10^{3}$ and $10^{4}$ respectively.  The masses are between 0.1 and
150 M$_{\odot}$ and are drawn from a \cite{KroupaEtAl93} initial mass function
with a low-mass slope of $\alpha_{l} = 1.3$, a high mass slope of $\alpha_{h}
= 2.3$ and a break mass of $M_{\rm   break} = 0.5$ M$_{\odot}$.

\begin{figure}
  \centering
  \includegraphics[clip=true,width=0.5\textwidth]{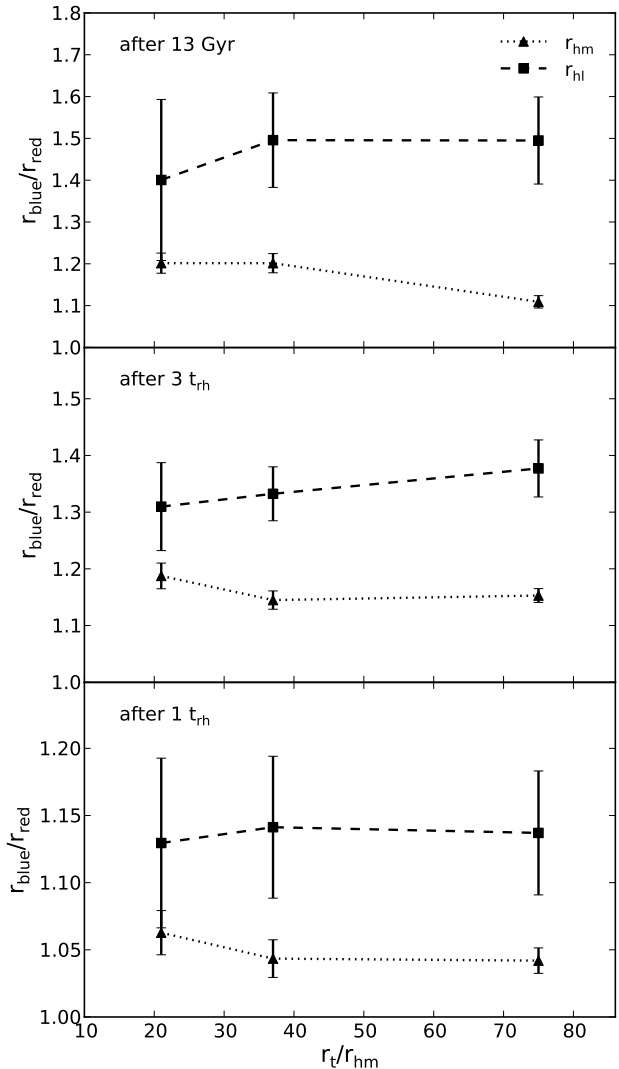}
  \caption{Ratio of $r_{hl}$ and $r_{hm}$ in blue globular clusters as
    compared to red globular clusters for clusters with different initial
    concentrations.  The top plot gives the ratio after 13 Gyr, the middle the
    ratio after three initial half-mass relaxation times, and the bottom the
    ratio after one initial half-mass relaxation time.  One sigma error bars
    are given.  Half-mass and half-light radii are clearly larger in blue
    clusters.\label{fig:HL-HM-13}}
\end{figure}

The simulations come in two metallicities, $Z = 0.02$ and $Z = 0.001$.
Throughout the paper I will refer to the $Z = 0.02$ clusters as the ``red''
clusters and the $Z = 0.001$ clusters as the ``blue'' clusters.  One
significant difference between the red and blue simulations is the treatment
of black holes (BHs).  Metal-poor stars lose less mass to line-driven winds
than do metal-rich stars and experience more matter fallback after supernovae
\citep{BelczynskiEtAl02, BelczynskiEtAl06}.  This produces both more and more
massive BHs in the blue clusters.  BHs also receive natal kicks drawn from a 
Maxwellian distribution with a peak at 190 km/s \citep{HansenPhinney97},
significantly higher than the escape velocity of a GC.  This kick is then
reduced according to the amount of fallback on to the BH during the supernova
\citep{BelczynskiEtAl02}.  Because the amount of fallback is larger in the
blue clusters, the BHs in these clusters receive a larger kick reduction and
are less likely to escape.  The combination of the larger BH masses and higher
BH retention rates leads to significant differences in the BH populations of
blue and red clusters.

\begin{figure}
  \centering
  \includegraphics[clip=true,width=0.5\textwidth]{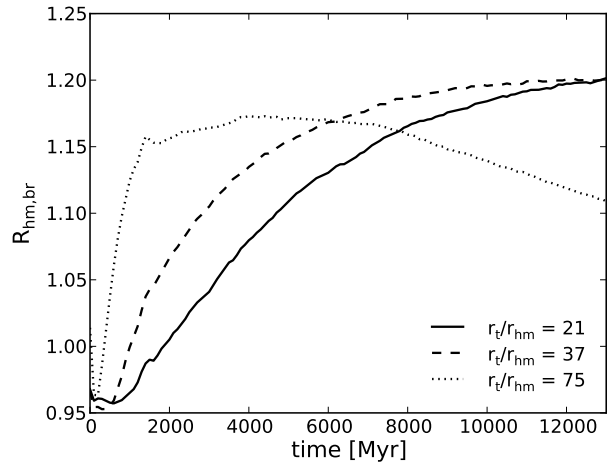}
  \caption{The evolution of $R_{hm, br}$ as a function of time for all three
    different initial concentrations.\label{fig:RhmTime}}
\end{figure}

The combination of initial concentration and metallicity give 6 different sets
of initial conditions, each of which is then independently realized 10 times
to constrain the stochasticity inherent in collisional stellar
dynamics.  All values in this paper are an average over all ten realizations
unless specifically noted otherwise. This yields a set of 60 simulations.  The
basic parameters are given in Table~\ref{tab:SimParams}.  The initial
half-mass relaxation times are calculated according to \cite{Spitzer87}:
\begin{equation}
  \label{eq:trh}
  t_{rh} = 0.138 \frac{N^{1/2} r_{hm}^{3/2}} {\langle m \rangle^{1/2} G^{1/2}
    \ln{\gamma N}},
\end{equation}
where $\langle m \rangle$ is the average mass of stars in the cluster.  I note
that each of these simulations takes only $\sim$ 1 day to run as opposed to
the months needed for direct $N$-body simulations.  It is this advantage in
speed that makes this investigation possible.  I have performed several
further simulations to investigate specific effects and these will be
described later.

\section{The Relative Sizes of Red and Blue GCs}
\label{sec:RelHM}

\begin{figure}
  \centering
  \includegraphics[clip=true,width=0.5\textwidth]{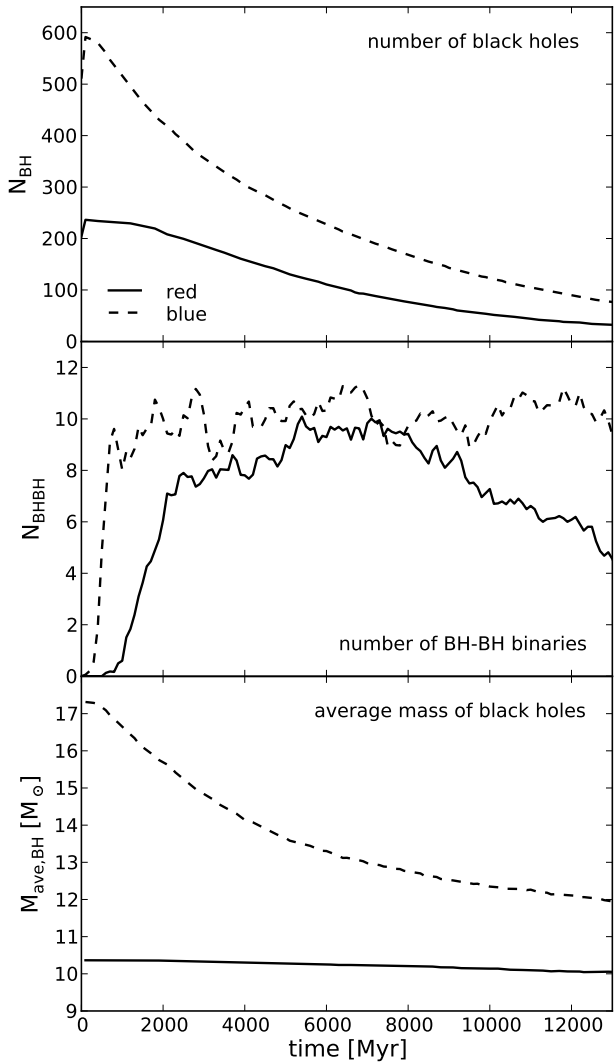}
  \caption{The evolution of the BH population in blue and red GCs with an
    initial concentration of $r_{t}/r_{hm} = 37$.  The top panel gives the
    number of single BHs, the middle panel the number of BH-BH binaries and
    the bottom panel the average mass of single BHs.\label{fig:BHpop}}
\end{figure}

In figure~\ref{fig:HL-HM-13} I show the ratios of $r_{hm}$ and $r_{hl}$
between blue and red GCs.  $r_{hl}$ is calculated by projecting the luminosity
of all stars into annular rings and summing the total luminosity in each ring
radially outwards until finding the radius containing half the total
luminosity.  The ratio of the value of $r_{hm}$ in blue over the value of
$r_{hm}$ in red clusters will hear after be referred to as $R_{hm, br}$ while
the ratio of the value of $r_{hl}$ in blue over the value of $r_{hl}$ in red
cluster will be referred to as $R_{hl, br}$.  It is clear that both $r_{hm}$
and $r_{hl}$ are enhanced in blue clusters.  After 13 Gyr the value of
$r_{hm}$ is $\sim 20\%$ greater in blue clusters, in rough agreement with the
value found after $\sim$ 250 Myr by \cite{Schulman12}.  By contrast the value
of $r_{hl}$ is some $\sim$ 40-50\% greater, a larger enhancement than observed
in real clusters.  The difference in $R_{hm, br}$ and $R_{hl, br}$ indicates
that the enhancement of $r_{hl}$ in blue GCs is not simply the result of a
difference in size.  That $R_{hl, br}$ is larger in my simulations than in the
observations suggests that my simulations over-predict the difference between
blue and red GCs.

\begin{figure}
  \centering
  \includegraphics[clip=true,width=0.5\textwidth]{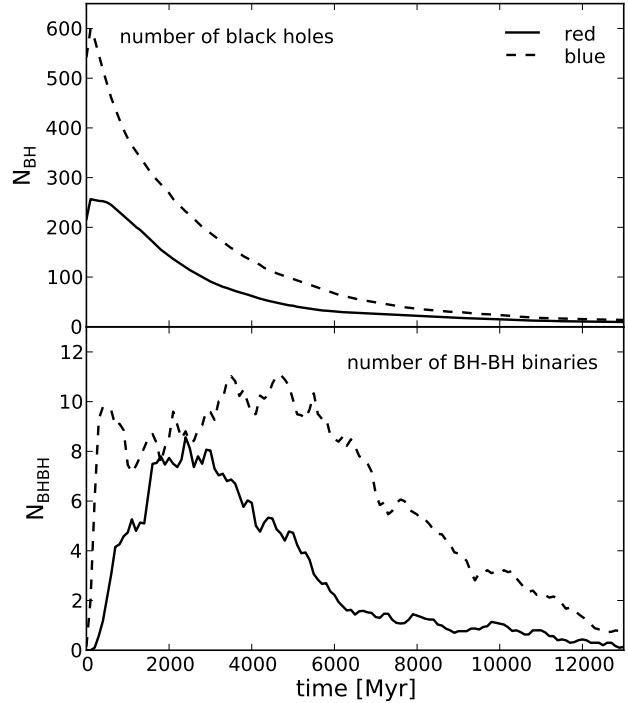}
  \caption{The BH and BH-BH binary population in clusters with $r_{t}/r_{hm} =
    75 $.  The top panel gives the number of single BHs while the bottom panel
    gives the number of BH-BH binaries.\label{fig:BHpop75}}
\end{figure}

The time evolution of $R_{hm, br}$, shown in figure~\ref{fig:RhmTime}, is
rather similar to the behaviour found by \cite{Schulman12} -- $r_{hm}$ starts
off smaller in the blue clusters but then grows rapidly so the blue clusters
become larger than the red ones within an initial relaxation time.
\cite{Schulman12} suggest that this pattern reflects stronger two-body
relaxation between the more massive stars in the blue clusters.  Both clusters 
experience strong initial mass-loss due to the rapid evolution of massive
stars and because this effect is stronger in the red clusters they initially
expand faster.  However, the more energetic two-body interactions between the
massive stars in the blue clusters generate more dynamical heat and cause them
to expand faster than the red clusters once the initial phase of rapid
mass-loss is complete.  Thus after the first few hundred Myrs the blue
clusters become larger than the red ones.

\begin{figure}
  \centering
  \includegraphics[clip=true,width=0.5\textwidth]{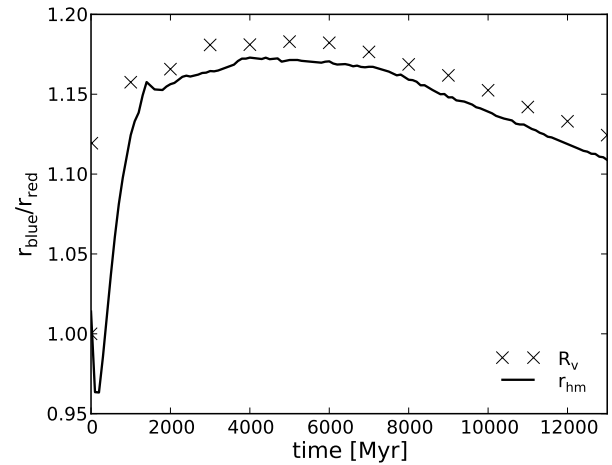}
  \caption{A comparison of ratios of $R_{v}$ and $r_{hm}$ between blue and red
    clusters with $r_{t}/r_{hm} = 75$.\label{fig:Rv75}}
\end{figure}

\cite{Schulman12} do not determine whether the dominant source of the
two-body heating in the blue clusters is the interactions between a large
number of stars, all of which are only slightly more massive than the stars
in the red clusters, or if a small population of very massive stars is
responsible for the effect.  A comparison of the time evolution of $R_{hm,
  br}$ in figure~\ref{fig:RhmTime} for different initial concentrations
provides a clue.  For the lowest density (dynamically youngest) clusters the
value of $R_{hm, br}$ slowly grows to $\sim 1.2$ and then remains constant
whereas in the densest (dynamically oldest) cluster $R_{hm, br}$ grows rapidly
but then begins to decline.  This indicates that the densest blue clusters
start to re-contract compared to the red clusters because the population of
stars responsible for the expansion has been lost.  Such behaviour suggests a
small population of massive objects that can pump a large amount of kinetic
energy into the blue clusters through interactions but are also scattered out 
of the cluster and depleted.

\begin{figure}
  \centering
  \includegraphics[clip=true,width=0.5\textwidth]{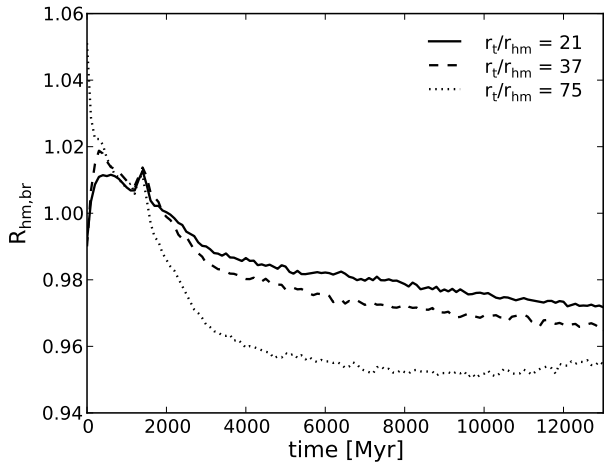}
  \caption{The same as for figure~\ref{fig:RhmTime} but for clusters where all
  BHs have been removed by giving them 1000 km/s natal
  kicks.\label{fig:RhmTimeNoBH}}
\end{figure}

BHs are excellent candidates for this population: after the first few 100
Myrs they are significantly more massive than the average mass in the cluster;
they are expected to mass-segregate to the core, become Spitzer-unstable
\citep{Spitzer87, SigurdssonHernquist93}, and interact strongly; they are also
few in number so they can be rapidly depleted by interactions.
\cite{MacKeyEtAl08} presented simulations that show a population of BHs can
lead to a significant growth in core radius compared to simulations without
BHs.  Figure~8 of the same paper suggests the same may be true for $r_{hm}$.
Specifically, \cite{MacKeyEtAl08} propose that the BHs mass-segregate to the
centres of the clusters where they strongly interact and form BH-BH binaries.
\cite{DowningEtAl10} show that this also occurs in my simulations.  These
binaries scatter single BHs from the cluster centre to its outer regions.
From there these BHs sink back to the centre of the cluster due to dynamical
friction.  This process does $m_{BH} | \phi |$ work on the cluster per BH
where $m_{BH}$ is the mass of the BH and $\phi$ is the gravitational potential
of the cluster.  This work dynamically heats the cluster and causes it to
expand.

\begin{figure}
  \centering
  \includegraphics[clip=true,width=0.5\textwidth]{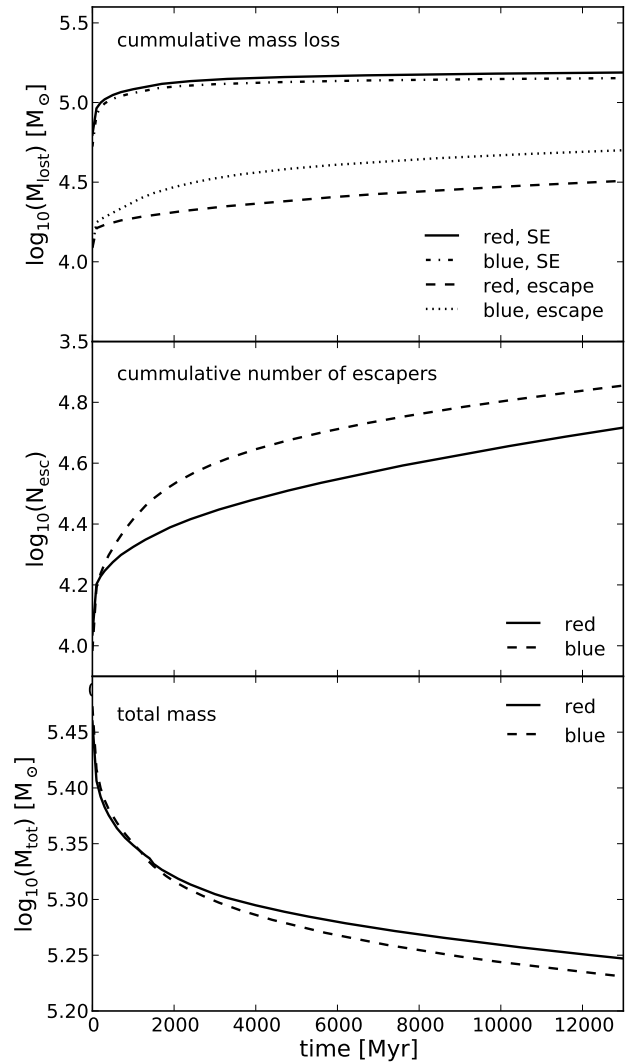}
  \caption{Mass-loss and number of escapers in blue vs. red clusters with BHs
    and with $r_{t}/r_{hm} = 37$.  The top panel gives the cumulative
    mass-loss due to stellar evolution and escapers, the middle the
    cumulative number of escapers and the bottom the total mass as a function
    of time.\label{fig:M-Mloss}}
\end{figure}

As discussed in \S~\ref{sec:Models}, there can be major differences between
the BH populations in my red and blue simulations.  I show the population of
BHs in the $r_{t}/r_{hm} = 37$ simulations in figure~\ref{fig:BHpop}.  I
choose to analyze this set of simulations because there are the least
extreme set of initial conditions and should be the most generally
representative.  Figure~\ref{fig:BHpop} confirms that the blue clusters indeed
have a larger population of BHs and that these BHs are significantly more
massive than those in the red clusters.  The blue clusters also produce
slightly larger numbers of BH-BH binaries slightly earlier than the red
clusters but the difference is not nearly as great.  This is in line with the
model of \cite{MacKeyEtAl08} where the BH-BH binaries do not interact 
directly with the rest of the cluster but only with other BHs.  It is these
scattered single BHs that are responsible for the dynamical heating of the
luminous stellar population.  Although the red clusters have not insubstantial
BH-BH binary populations compared to the blue clusters, they do not have a
sufficient number of sufficiently massive single BHs to heat the rest of the
cluster and cause it to expand.

\begin{figure*}
  \centering
  \includegraphics[clip=true,width=\textwidth]{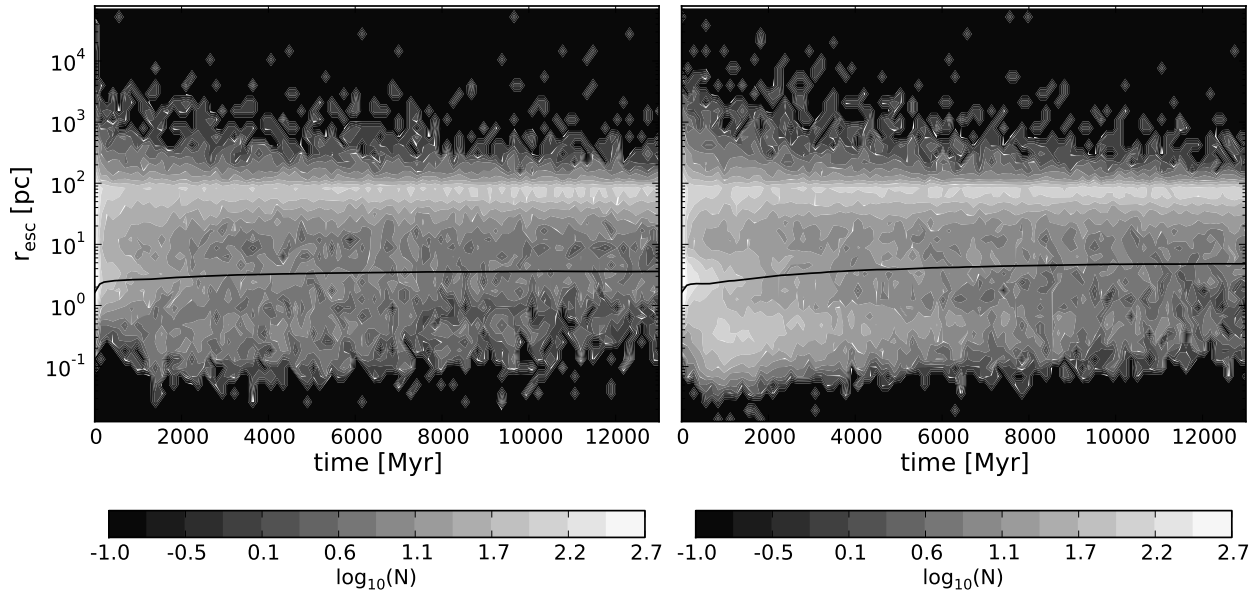}
  \caption{Position of last interaction before escape relative to the cluster
    centre for escapers from a single red (left) and single blue (right)
    cluster with $r_{t}/r_{hm} = 37$ as a function of time.  The black line is
    the 10\% Lagrangian radius.\label{fig:rEsc}}
\end{figure*}

The BH hypothesis also explains the evolution of the most concentrated
clusters.  Figure~\ref{fig:BHpop75} shows that the both BHs and BH-BH binaries
are rapidly depleted in the $r_{t}/r_{hm} = 75$ clusters.  Consequently
these clusters loses this source of dynamical heat.  The Monte Carlo models
remain in virial equilibrium to within a fraction of a per cent by
construction so this loss of heat must be compensated for by a change in the
structure of the cluster.  The virial radius,
\begin{equation}
R_{v} = \frac{GM^{2}}{2K}, \qquad \left( K = \frac{1}{2}M\sigma^{2} \right)
\end{equation}
where $M$ is the total mass of the cluster, $\sigma$ the velocity dispersion
and $K$ the kinetic energy, is a good measure of this change.
Figure~\ref{fig:Rv75} shows that the ratio of $R_{v}$ between the blue and red
clusters increases while there are many BHs and BH-BH binaries and then
decreases when these objects are depleted.  Indeed the drop in the ratio of
$R_{v}$ and a sharp drop in the number of BH-BH binaries both occur at $\sim
6$ Gyr.  The ratio of $R_{v}$ also tracks the evolution of $R_{hm, br}$ very
closely.  This strongly supports the case that heating by BHs is responsible
for the size differences between these clusters.

\begin{figure}
  \centering
  \includegraphics[clip=true,width=0.5\textwidth]{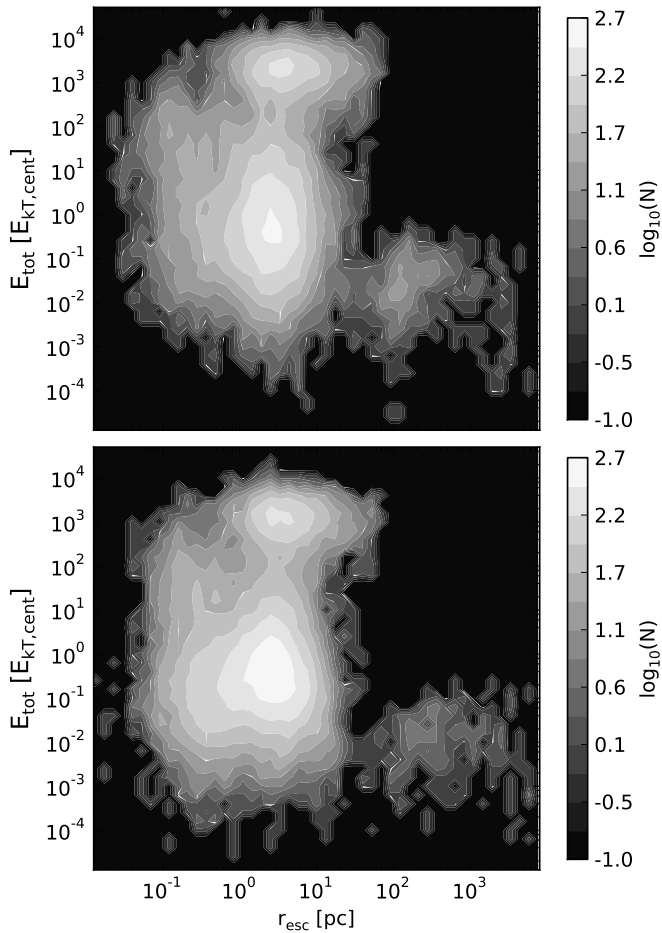}
  \caption{The total energy of escaping stars as a function of the position of
  their last interaction before escape for a single red (top) and single blue
  (bottom) cluster with $r_{t}/r_{hm} = 37$.  The total energy of each star is
  given in units of the core kinetic energy of the cluster.  The low-energy
  peak is primarily ejections due to two-body relaxation while the high-energy
  peak is ejections due to interactions with binaries.\label{fig:eEsc}}
\end{figure}

To confirm that the difference in $r_{hm}$ between my blue and red clusters is
a result of the different BH populations I have run a set of simulations
without BHs.  For each set of initial conditions listed in
table~\ref{tab:SimParams} I have performed an additional five simulations,
identical up to and including the random seed, but where all BHs are given a
kick of 1000 km/s upon formation.  This instantaneously removes all black
holes while allowing the cluster to continue its evolution.  Values of $R_{hm,
  br}$ for the clusters without BHs are given in figure~\ref{fig:RhmTimeNoBH}
and can be compared to the results for clusters with BHs given in
figure~\ref{fig:RhmTime}.  The results are dramatic -- the enhancement in
$r_{mh}$ of blue clusters completely disappears!  Indeed the trend observed in
figure~\ref{fig:RhmTime} is reversed: the blue clusters are slightly larger at
early times -- probably due to the fact that BHs are more numerous and massive
in these clusters, resulting in more mass loss when they are ejected -- and
slightly smaller at late times.  The effect in figure~\ref{fig:RhmTimeNoBH} is
only $\sim 2-4 \%$, much smaller than the $\sim 20\%$ differences observed in
figure~\ref{fig:RhmTime}.  Because the only difference between these two sets
of simulations is the BH population, this proves that the enhancement in
$r_{hm}$ observed in my first set of blue clusters is due to the dynamical
activity of BHs.

\begin{figure}
  \centering
  \includegraphics[clip=true,width=0.5\textwidth]{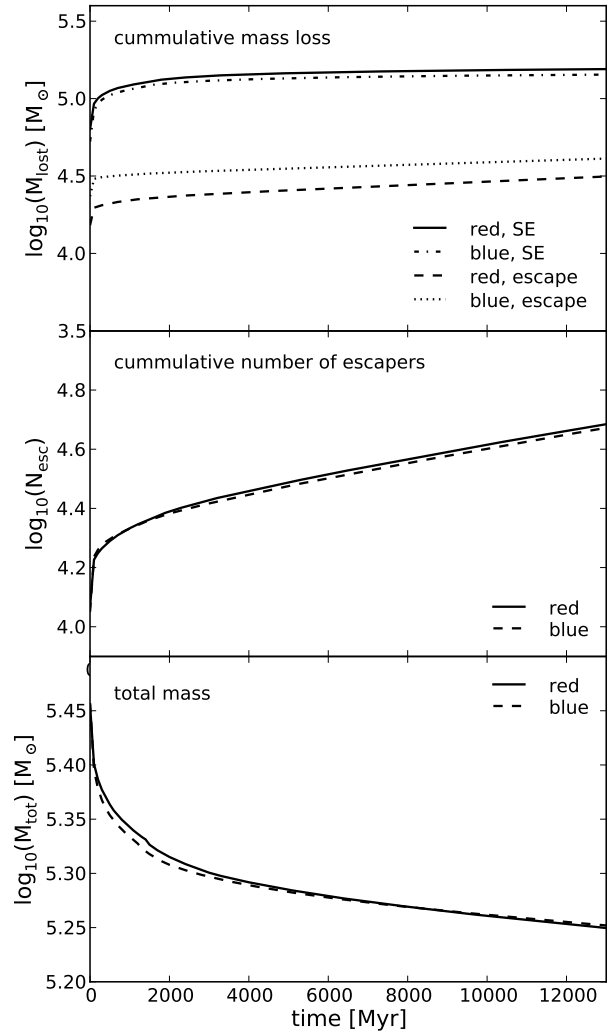}
  \caption{The same as for figure~\ref{fig:M-Mloss} but for clusters without
    BHs.\label{fig:M-Mloss-NoBH}}
\end{figure}

One effect I observe in my simulations with BHs that was not extensively
discussed by \cite{MacKeyEtAl08} is a difference in the total escape rate of
stars from the blue and red clusters (\cite{MacKeyEtAl08} examine only the
escape rate of BHs).  In figure~\ref{fig:M-Mloss} I show that the blue
clusters lose more stars, both by number and by mass, to escape than do the
red clusters.  By contrast the red clusters lose slightly more mass to stellar
evolution than do the blue clusters.  The difference in mass lost to escapers
is larger and the blue clusters lose more mass overall than do the red
clusters.  Figure~\ref{fig:rEsc} shows that the majority of these additional
escapers come from the central regions of the cluster, which suggests they are
stars that have gained additional energy through interactions with the BH
sub-system.  This additional mass-loss from the cluster centre will cause the
inner Lagrangian radii to expand compared to the overall radius of the cluster
and enhance the expansion of $r_{hm}$.  Figure~\ref{fig:eEsc} gives the total
energy of each escaping star ($E_{\star} = K_{\star} + m_{\star}\phi$ where
$K_{\star}$ is the kinetic energy of the star and $m_{\star}$ is its mass) as
a function of the position of its last interaction before escaping from the
cluster.  Most of the additional escapers in the inner region do not have
particularly high energies.  Therefore they are unlikely to have been ejected
in an interaction with a binary.  Rather they have gained their energy due to
strong two-body relaxation in the inner regions of the cluster.  This is fully
consistent with the results of \cite{MacKeyEtAl08} who found that the BH-BH
binaries do not interact with other cluster stars directly but only with the
single BHs.  It is two-body heating from these single BHs that produce the
expansion of the cluster and the higher escape rate.

\begin{figure}
  \centering
  \includegraphics[clip=true,width=0.5\textwidth]{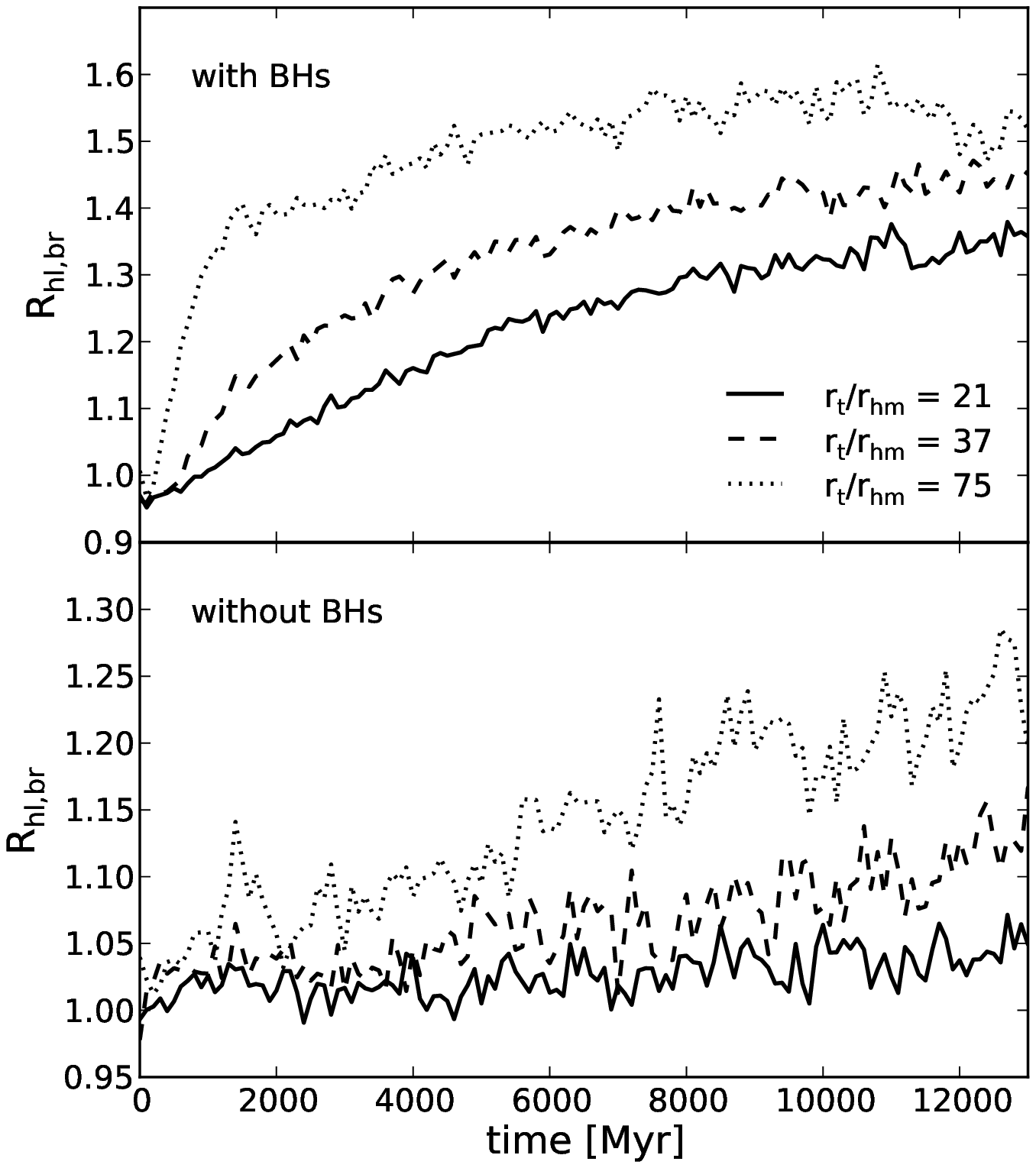}
  \caption{The ratio of $r_{hl}$ between blue and red clusters for all initial
    concentrations with and without BHs.  The top panel gives the values for
    clusters with BHs while the bottom panel gives the values for clusters
    without BHs.\label{fig:RhlTime}}
\end{figure}

In figure~\ref{fig:M-Mloss-NoBH} I compare the mass-loss and escaper rates in
blue and red clusters without BHs.  The difference in total number of escapers
almost completely disappears.  There is still a difference between the mass
lost due to escapers but the majority of this occurs early and can be
attributed to the instantaneous ejection of BHs from the clusters.  Because
the blue clusters form more BHs and these BHs are more massive, it follows
that the blue clusters will lose more mass when they are ejected.  This may
also explain the slightly larger values of $r_{hm}$ in blue clusters without
BHs at early times.  This difference has little effect on the total mass of
the clusters and, unlike the case with BHs, the blue and red clusters have
approximately the same mass at 13 Gyrs of age.

\section{The half-light radii of red and blue GCs}
\label{sec:Rhl}

\begin{figure}
  \centering
  \includegraphics[clip=true,width=0.5\textwidth]{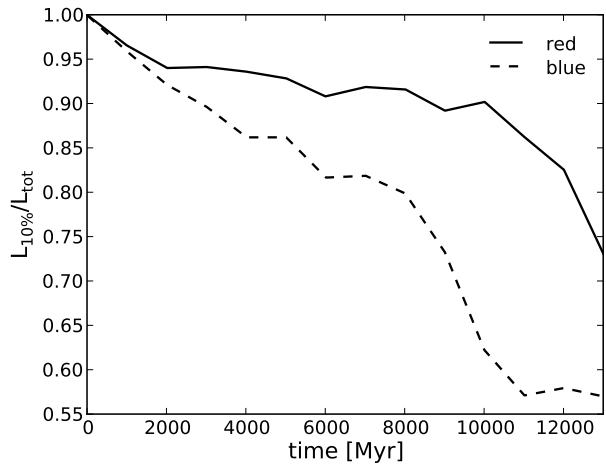}
  \caption{The fraction of the total luminosity contained within the most
    massive 10\% of stars for blue and red clusters without BHs and with
    $r_{t}/r_{hm} = 37$.\label{fig:FracLum}}
\end{figure}

Figure~\ref{fig:HL-HM-13} shows that the value of $R_{hl, br}$ is larger both
than $R_{hm, br}$ and than the value that is observed in real GCs.  Therefore
it is clear from these simulations that $R_{hl, br}$ is not necessarily a good
predictor of $R_{hm, br}$.  To understand the difference between $R_{hl, br}$
and $R_{hm, br}$ I compare the time evolution of $R_{hl, br}$ in clusters with
and without BH in figure~\ref{fig:RhlTime}.  The half-light radii are
clearly larger in blue clusters for all sets of initial conditions, regardless
of whether $r_{hm}$ is larger or not.  The enhancement in $r_{hl}$ in the
clusters without BHs is smaller but is actually in much better agreement with
the observed enhancement of $\sim 20\%$, at least for the two most initially
concentrated clusters.  The difference between $r_{hm}$ in the blue and red
clusters with BHs contributes to the difference in $r_{hl}$ but clearly there
are other processes at work.

\begin{figure}
  \centering
  \includegraphics[clip=true,width=0.5\textwidth]{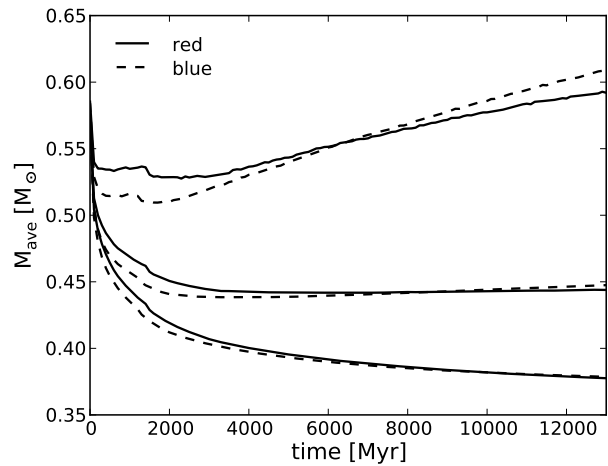}
  \caption{From top to bottom: the average mass within the 10\%, 50\% and
    100\% Lagrangian radii as a function of time for red and blue clusters
    without BHs and with $r_{t}/r_{hm} = 37$.\label{fig:Mseg}}
\end{figure}

In figure~\ref{fig:FracLum} I compare the fraction of the total luminosity in
the most massive 10\% of stars for the blue and red clusters while in
figure~\ref{fig:Mseg} I compare the mass-segregation in these clusters using
the average masses within selected Lagrangian radii.  The red clusters have a
steeper luminosity function with a significantly larger proportion of their
luminosity concentrated in their most massive stars.  The average stellar
mass is higher for the inner Lagrangian radii in both blue and red clusters so
both sets of clusters are mass-segregated.  For a given Lagrangian radius,
however, the average mass is similar in both the blue and the red clusters.
Taking figures~\ref{fig:FracLum} and figure~\ref{fig:Mseg} together, this
means that the red clusters will have a larger fraction of their luminosity
located within their innermost Lagrangian radii than will the blue clusters
and thus they will have a smaller value of $r_{hl}$ compared to $r_{hm}$.
Consequently the value of $R_{hl, br}$ will be larger than the value of
$R_{hm, br}$.  This explains why $R_{hl, br} > 1$ in clusters of the same size
according to $r_{hm}$.  The effect will be further enhanced if the blue
clusters are also larger than the red clusters and explains the greater value
of $R_{hl, br}$ in clusters with BHs.  However it is not necessary for a
cluster with a larger value of $r_{hl}$ to have a larger value of $r_{hm}$.
This suggests an additional corollary: $r_{hl}$ cannot be used to measure the
relative sizes of GCs, at least not in a straightforward way.

\section{Discussion}
\label{sec:discussion}

My simulations show that blue GCs can be larger than red GCs but only if they
have a substantial population of BHs.  Blue and red clusters without BHs will
be roughly the same size but the red clusters will still have smaller values
of $r_{hl}$ due to their steeper luminosity functions.  I find that the
enhancement in $r_{hl}$ for blue clusters with BHs is significantly larger
than that observed value.  In the clusters without BHs the enhancement in
$r_{hl}$ roughly matches the observations, at least for the clusters with the
highest initial concentrations.  This finding supports the results of
\cite{Jordan04} -- that the difference in half-light radii between blue and
red GCs is the result of different luminosity functions in clusters with
similar sizes -- rather than the results of \cite{Schulman12}.  It may also
indicate that there is little difference between the BH population in blue and
red GCs.  This will have consequences for understanding effect of
metallicity on the late-time evolution and supernovae of massive stars.

The simulations of \cite{Schulman12} used the initial mass function of
\cite{Kroupa01}, which has a high-mass slope of $\sim -2.3$, and used a
maximum Mass of $50$ M$_{\odot}$ (Glebbeek, private communication).  Such a
mass function will yield few BHs so a question remains as to why my
simulations with BHs behave in a qualitatively similar way to those of
\cite{Schulman12} while the ones without BHs do not.  I speculate that the
similarity is a result of the differing relationship between the stellar
evolution and the relaxation timescales.  In my simulations $t_{rh}$ is
significantly longer than the evolution timescale for massive stars, even in
the most concentrated clusters.  Thus by the time relaxation becomes important
for the structure of the cluster the only remaining population of stars
significantly more massive than average are the BHs.  By contrast the
relaxation timescales of the simulations of \cite{Schulman12} are $\sim$10-50
Myr, shorter than the evolution timescales of moderately massive stars.
Therefore these stars remain when two-body relaxation starts to drive cluster
evolution.  Furthermore, the short relaxation time will allow these stars to
mass-segregate before they lose a significant amount of their mass to winds
and nuclear burning.  Interactions between these stars will be able to act as a
heat source in a similar way to the BH sub-system in my simulations.  I
hypothesize that small clusters, with relaxation times shorter than the
evolution timescale for massive stars, have a variety of objects massive
enough to drive cluster expansion through two-body heating while in GCs, where
the relaxation timescale is much longer than the stellar evolution timescale,
BHs are the only remaining objects massive enough to produce this effect.

\begin{figure}
  \centering
  \includegraphics[clip=true,width=0.5\textwidth]{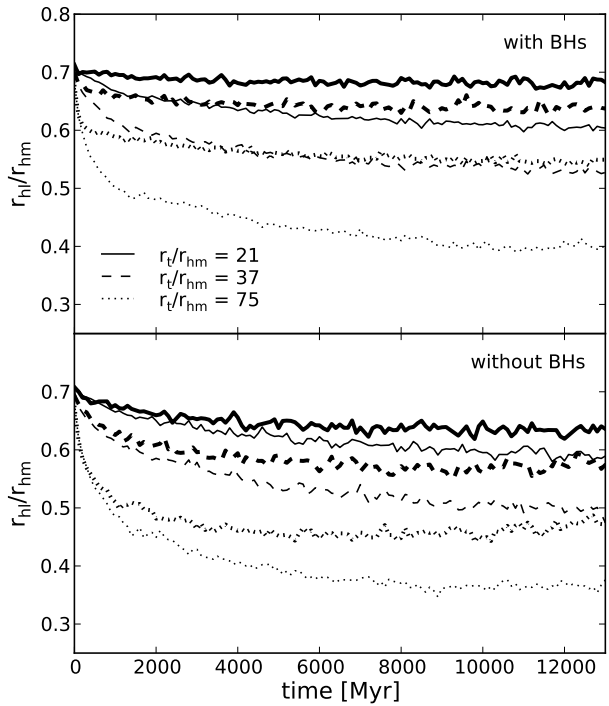}
  \caption{The ratio of $r_{hl}$ to $r_{hm}$ for all simulated GCs.  Thin
    lines are red GCs while thick lines are blue GCs.  The top panel gives the
    values for GCs with BHs while the bottom panel gives values for GCs
    without BHs.\label{fig:RhlRhm}}
\end{figure}

It is also apparent that the value of $R_{hl, br}$ is not the same as that of
$R_{hm, br}$ and thus it cannot be used as a direct estimate of the relative
sizes of GCs.  Indeed a comparison of figures~\ref{fig:RhmTimeNoBH}
and~\ref{fig:RhlTime} shows that a cluster can have a smaller value of
$r_{hm}$ than another cluster but still have a larger value of $r_{hl}$.  In
figure~\ref{fig:RhlRhm} I show the ratio of $r_{hl}$ to $r_{hm}$ for all of my
cluster models.  The ratio varies considerably both with time and with the
initial conditions used.  The ratio does seem to attain a fairly stable value
for each individual cluster, indicating that $r_{hl}$ is not completely
independent of $r_{hm}$.  Since this value always lies somewhere between 0.35
and 0.7 it is, in principle, possible to use $r_{hl}$ to estimate the value of
$r_{hm}$ to within a factor of $\sim 2$.  It is not possible, however, to use
$R_{hl, br}$ to estimate the relative size of two clusters unless (at minimum)
their metallicities, initial concentrations, and ages are known.

\section{Conclusions}
\label{sec:conclusions}

I have analyzed a suite of Monte Carlo globular cluster simulations to
determine the origin of the difference in half-light radii between blue and
red GCs.   I find that, provided they are mass-segregated, the value of
$r_{hl}$ is larger in blue than in red GCs simply due to differences in the
luminosity function between metal-poor and metal rich stellar populations.
Depending on the initial conditions this effect is sufficient to explain the
observed differences between $r_{hl}$ in blue and red GCs.  I find that it is
also possible for blue clusters to be physically larger than red clusters as
measured by their half-mass radii.  However this only occurs if there are
significant differences in the number and properties of BHs between metal-poor
and metal-rich stellar populations.  A difference in $r_{hm}$ can certainly
enhance the difference in $r_{hl}$ but it is not a necessary condition for
such a difference to exist.  Furthermore, the enhancement in $r_{hl}$ in blue
GCs when a there is significant difference in the BH population is larger
than observed.  This leads me to conclude that blue and red GCs probably do
not have significant differences in their BH populations but further
simulations and more careful comparisons with observations will be necessary
to confirm this.  Further simulations are also necessary to determine if it is
number or mass that determines whether BHs can cause a GC to expand.  Finally,
I find that $R_{hl, br}$ does not directly predict $R_{hm, br}$ so differences
in $r_{hl}$ cannot be used to infer differences in size between clusters of
different metallicities.

\section*{Acknowledgements}

Funding for this project was supplied by VESF grant EGO-DIR-50-2010 and the
simulations were carried out at the High-Performance Computing Centre Stuttgart
(HLRS) through the Baden-W\"u{}rttemberg grid (bwgrid) and the Astrogrid-D and
D-Grid initiatives.  I would like to thank Christoph Olczak at the ARI and
Mirek Giersz at the Copernicus Institute in Warsaw for many insightful
discussions and suggestions during the course of this work.



\label{lastpage}

\end{document}